\documentclass[]{article}

\usepackage{amssymb}

\begin{document}

\hfill {\footnotesize WM-02-113} 

\vskip 4.5ex

\centerline{\bf EXCITED BARYON PRODUCTION AND DECAYS\footnote{Invited talk at the workshop on The Phenomenology of Large $N_c$ QCD, held at Arizona State University, Tempe, AZ, USA, 9-11 January 2002.
}}

\vskip 4ex

\centerline{CARL E. CARLSON}

\begin{centering}

{\em Nuclear and Particle Theory Group, Department of Physics,

College of William and Mary, Williamsburg, VA 23187-8795, USA

E-mail: carlson@physics.wm.edu}

\end{centering}

\vskip 1ex

{\baselineskip 10pt

\begin{quote}
{
\footnotesize
We consider decays of the lowest-lying positive parity ({\bf 56}-plet)
and negative parity ({\bf 70}-plet) excited  baryons.  For the {\bf
70}-plet, we include both single-quark and two-quark decay operators,
and find, somewhat mysteriously, that the two-quark operators are not
phenomenologically important.  Studies of decays {\bf 70} $\rightarrow
\Delta + \gamma$ may strengthen or vitiate this observation.  For the
{\bf 56}-plet decays, now using only the single-quark operator, we can
predict many strong decays after fitting parameters on the assumption
that the Roper is a 3$q$ state. Comparison of these predictions to
experiment can verify the structure of the {\bf 56}-plet.  As a
sidelight, we show a large $N_c$ derivation of the old
G\"ursey-Radicati mass formula.
}
\end{quote}

}

\vskip 3 ex


{\noindent \bf 1. Introduction}


\vskip 2ex

\noindent I should note that I have written on large $N_c$ topics in collaboration
with the chief organizer of this conference and with the chairman of this
session.  Those works~\cite{largen1} concerned mass operators for
excited baryons using large $N_c$ to segregate terms by size, and a
continuation of the mass analysis work is reported elsewhere in this
conference~\cite{schat}.  This talk is not about that work.

This talk will discuss the decays of excited baryons, in particular strong
decays of the excited positive parity states~\cite{largen3} and
electromagnetic decays of the negative parity states~\cite{largen2}
(work from a similar viewpoint on strong decays of the latter are reported
in~\cite{carone1994}). Large
$N_c$ ideas have contributed directly to the analysis we will report on by
categorizing operators by order in $N_c$. In addition, we will show a
large
$N_c$ derivation of an old mass formula~\cite{GR}. We need this mass
formula because not all interesting states are yet
found experimentally, and masses and phase space are crucial to
calculating decay rates. It also happens that the large $N_c$ derivation
of the mass formula is new and interesting.

As mentioned, we have looked at two sets of states. One set is the
lowest-lying positive parity excited states of the baryons.  Baryons in
the ground state form a 56-dimensional representation of $SU(6)$.  The
excited positive parity states also form a {\bf 56}-plet, the {\bf
56}$'$.  The other set is the non-strange members of the lowest lying
negative parity baryon excitations, which are members of a {\bf 70}-plet
in $SU(6)$.  

Regarding the {\bf 70}-plet and the electromagnetic decays, the old and
well known~\cite{algebraic,close} treatment identifies three independent
operators that involve a single quark.  The 19 possible decay
amplitudes have been measured~\cite{pdg}, and one may try to fit them in
terms of matrix elements of the three single quark operators, with the
overall coefficient for each operator fit to the data.  The resulting
fit is not awful, although with a $\chi^2$ about 4 per independent data
point, one might hope for better.  There are additional operators that
involve other quarks as well, and one should not {\it a priori} neglect
them in a low momentum transfer strong interaction process.  We
examined the 2-body operators~\cite{largen2} and found many that are
{\it not}, contrary to our hope, suppressed by large  $N_c$ arguments.
Regarding how much including them improved the fit to the data, we will
let the reader find in Section~2.  A hint may be found by scanning the
conclusions in Section~5.

Preceding the decay discussion for the {\bf 56}$'$ will be a seemingly
tangential discussion of the G\"ursey-Radicati mass formula~\cite{GR}. 
This is a 4-parameter mass formula for a {\bf 56}-plet of $SU(6)$
(which has 8 independent masses, not counting isospin splittings).  It
applies to the ground state baryons, but also to the positive parity
excited {\bf 56}$'$.  There are two reasons for including the
G\"ursey-Radicati mass formula discussion.  One is that the original
1964 derivation is not to the modern taste, whereas a derivation using
large $N_c$ ideas is both clear and illustrative of how large $N_c$
ideas let us estimate the sizes of terms~\cite{manrev}, allowing us to
keep the few largest mass operators to produce a mass relation, and to
use the largest omitted term to estimate its error.  The second reason is that not all the {\bf 56}$'$ states have been experimentally found, and we need the mass formula to get the masses of the missing states. The mass formula is discussed in Section~3.

Were one to select a prime motive for the {\bf 56}$'$ strong decay
discussion, it would be to help determine if the Roper be dominantly an
ordinary $3q$ baryon state~\cite{isgurkarl1,glozmanriska}, or something more
exotic such as a hybrid $q^3 g$ state~\cite{hybridroper}, or something not
really a baryon state at all, but just an enhancement in some partial wave in a
scattering process~\cite{dynamicalroper}.

We shall be treating the Roper as if it is a $3q$ member of a 
{\bf 56}-plet of $3q$ states, and calculate decay rates based on assuming
the dominance of the (only possible) single quark decay operator~\cite{largen2},
an assumption that follows from the more detailed analysis made in the
context of the negative parity {\bf 70}-plet.  The details of the
analysis and predictions for many decays are presented in
Section~4.

Conclusions, again, are in Section~5.

\vskip 3ex


{\noindent \bf 2. Radiative Decays  of the {\bf 70}}

\vskip 2ex


\noindent The work described here follows upon the large $N_c$ revival of the middle
1990's~\cite{djm1,djm2,cgo,ml}.  Most applications were to the ground
state~\cite{manrev}, but the excited baryons were not entirely
neglected~\cite{goity,largen4}.  In particular, strong decays of the
{\bf70}-plet  were studied~\cite{carone1994} in a manner one can recognize from
the study of the electromagnetic decays that follows (though familiarity with
the earlier work is not assumed in what is written here).

One does know~\cite{close} that the {70-plet} of $SU(6)$ has one unit of
orbital angular momentum ($L=1$) and negative  parity.  It has 20 states
that are not strange, and if one lumps the isospin partners
together, one is left with 7 independent states.  

Two of the states are $\Delta$'s, characterized by $I=3/2$, and denoted
$\Delta_J$,
\begin{equation}
\Delta_{1/2,\, 3/2} = \left|I= \frac{3}{2},S=\frac{1}{2},L=1,
   J=\frac{1}{2},\frac{3}{2}  
                            \right\rangle   \ .
\end{equation}

\noindent Five of the states are nucleons, with $I=1/2$, and distinguished by
whether the core quark spins are combined to give $S = 1/2$ or $3/2$,

\begin{eqnarray}
N_{1/2, \, 3/2} &=& \left| I= \frac{1}{2},S=\frac{1}{2},L=1,
   J=\frac{1}{2},\frac{3}{2}  
                            \right\rangle   \ ,
          \nonumber \\
N'_{1/2, \, 3/2, \, 5/2} &=& \left|I= \frac{1}{2},S=\frac{3}{2},L=1,
   J=\frac{1}{2},\frac{3}{2},\frac{5}{2}  
                            \right\rangle   \ .
\end{eqnarray}

\noindent Since nucleon states with the same $J$  can mix, the physical
$N_{1/2, \, 3/2}$ states are the linear combinations
\begin{eqnarray}
\left[\begin{array}{c} N(1535) \\ N(1650) \end{array} \right] &=&
\left[\begin{array}{cc}  \cos\theta_{N1} & \sin\theta_{N1} \\
                       -\sin\theta_{N1} & \cos\theta_{N1}
\end{array}\right]
\left[\begin{array}{c} N_{1/2} \\ N'_{1/2}\end{array} \right]  \ ,
         \nonumber \\\
\left[\begin{array}{c} N(1520) \\ N(1700) \end{array} \right] &=&
\left[\begin{array}{cc}  \cos\theta_{N3} & \sin\theta_{N3} \\
                       -\sin\theta_{N3} & \cos\theta_{N3}
\end{array}\right]
\left[\begin{array}{c} N_{3/2} \\ N'_{3/2}\end{array} \right]  \ .
\end{eqnarray}

\noindent Thus the non-strange sector of the {\bf 70}-plet has 2 mixing
angles and 7 masses.

Consider decays in the $N^*$ rest frame (where $N^*$ is a member of the 
{\bf 70}-plet).
Using the outgoing photon direction as the helicity direction for the
$N^*$, one has
\begin{equation}
\lambda = \lambda_\gamma - \lambda_N = \pm \frac{3}{2}, \pm
\frac{1}{2}    \ ;
\end{equation}

\noindent giving the helicity $\lambda$ of the $N^*$ also
specifies the helicity $\lambda_\gamma$ of the photon and of the ground
state exiting nucleon $\lambda_N$.

The decay amplitudes $A_\lambda$ are matrix elements~\cite{old} of decay
operators
${\cal O}_i$
\begin{equation}
A_\lambda = \left\langle N,\gamma \left| 
             \sum {\cal O}_i  
             \right|  N^*, \lambda \right \rangle \ ,
\end{equation}

\noindent and by parity, $A_{3/2}$ and $A_{1/2}$ suffice~\cite{old}.

We have to consider how to make the operators ${\cal
O}_i$.  We can use the  \medskip

\begin{tabular}{ll}
\quad quark charge matrix & $Q$,  \\
\quad photon polarization & $\vec A \propto \vec\xi$, \\
\quad photon momentum & $\vec k$ \quad (or derivative $\vec\nabla$),  \\
\quad P-wave quark polarization  & $\vec\varepsilon$,   \\
\quad spin operator for quark  &  $\vec\sigma$.  \
\end{tabular}

\medskip

\noindent With these ingredients, there are 3 one-body operators ${\cal
O}_i$,
\begin{eqnarray}
&& a_1 Q_* \ (\vec{\varepsilon} \cdot \vec{A}),
                   \nonumber 
                   \\
&& b_1 Q_* \ (\vec{\varepsilon} \cdot \vec k)
\ (\vec{\sigma}_*\cdot\vec k
\times \vec{A}) ,
                   \nonumber
                   \\
&& b_2 Q_* \ (\vec{\sigma}_*\cdot \vec k) \ (\vec{\varepsilon}
\cdot \vec k \times \vec{A}),
\end{eqnarray}

\noindent which are equivalent to the textbook~\cite{close} SU(6)
one-body operators.  The star means an operator acting on the initial
P-wave quark.

Examples of two-body operators include
\begin{eqnarray}
&& \frac{c_3}{N_c} 
\Bigl(\sum_{\alpha \neq *} Q_\alpha \vec{\sigma}_\alpha \Bigr)\cdot
\vec{\sigma}_*
(\vec{\varepsilon} \cdot \vec{A}) ,
                   \nonumber 
                   \\
&& \frac{d_2}{N_c}
\Bigl(\sum_{\alpha \neq *} Q_\alpha \vec{\sigma}_\alpha \Bigr) \cdot
\vec k  \
(\vec{\varepsilon} \cdot \vec k) \ 
( \vec{\sigma}_* \cdot \vec{A}),
                   \nonumber 
                   \\
&& \frac{d_3}{N_c}
\Bigl(\sum_{\alpha \neq *} Q_\alpha \vec{\sigma}_\alpha \Bigr) \cdot
\vec k  \
(\vec{\sigma}_* \cdot \vec k) \ 
( \vec{\varepsilon} \cdot \vec{A}),
                   \nonumber 
                   \\
&& \frac{d_4}{N_c} \Biggl[\Bigl(\sum_{\alpha \neq *} Q_\alpha
\vec{\sigma}_\alpha \Bigr)
\times \vec{\sigma}_* \cdot
\vec k \, \Biggr] (\vec{\varepsilon} \cdot \vec k
\times \vec{A}) .
\end{eqnarray}

\noindent There are 8 leading-order (in $N_c$) 2-body operators, which are
listed in~\cite{largen2}.  The explicit factors of $N_c$ are included,
so that one may expect that the $a_i$, $b_i$, $c_i$, and $d_i$ are all
$O(N_c^0)$.

But now comes an important observation.  The 2-body operators have
matrix elements $\propto N_c$ (since they get one contribution from each
on the non-excited quarks in the baryon).  Hence their contributions to
the decay rates are not suppressed by large $N_c$ arguments as $N_c
\rightarrow
\infty$, compared to the 1-body operators.

Following this statement comes a surprise.  The   2-body operators are
empirically  unimportant.

Fitting data with 1-body operators alone gives $\chi^2=52.97$ (with
mixing angles fixed at $\theta_{N1}=0.61$ and $\theta_{N3}=3.04$,
which were obtained from an earlier analysis of the hadronic decays of
the {\bf 70}-plet~\cite{carone1994}; we can also do the fit letting
the mixing angles be fit using the electromagnetic data only, and get
the same mixing angles within the uncertainty limits).

Now include one 2-body operator with the
three one-body operators. The resulting $\chi^2$ depends on which
2-body operator is chosen, as tabulated here:

\medskip

\noindent
\begin{tabular}{ccccccccc} \hline
2-body op.   &  $c_1$ & $c_2$ & $c_3$ & $c_4$ 
              & $d_1$ & $d_2$ & $d_3$ & $d_4$ \\ \hline 
$\chi^2$      &  52.85 &  52.04 &  39.21 &  52.91    
              &  48.38 &  52.81 &  52.23 & 52.59  \\ \hline
\end{tabular}

\medskip

\noindent One does best using $c_3$, and the result including all 8
two-body operators simultaneously is not notably better than using $c_3$
alone.  Including the 2-body operators does not significantly improve the
fit to the data.

The dynamics will have to be understood better if we are
to understand why the the 2-body operators enter with small effect.
Large $N_c$ is not violated, in that no coefficient is forced to be
larger than expected, but it does appear that the 2-body terms can have
coefficients that are smaller that they have to be from large $N_c$ ideas.

One should say that there is an experimental opportunity here. Current
data on $N^*$ ``decays" really come from pion photoproduction, using
time-reversal invariance.  One then understands why all the decays that
have been discussed have a ground-state nucleon in the final state.  
One can anticipate that modern machines
(for example, Jefferson Lab) can produce enough $N^*$'s to see radiative
decays directly. That means that one can also think about decays into
$\Delta\gamma$ final states.

There are 24 (non-isospin related) amplitudes for the decays  
\begin{equation}
N^* {\rm \ or \ } \Delta^* \rightarrow \Delta\gamma    \ ;
\end{equation}

\noindent the subscripts on the amplitudes $A_\lambda$ can now run from
$-1/2$ to $5/2$ (with parity invariance allowing us to fix
$\lambda_\gamma = 1$).

The 1-body operators and their coefficients are the same as for the decays
into $N \gamma$; only the evaluation of the matrix elements changes.
Hence each of the 24 decays is predicted~\cite{largen3,gilman1974}.  We can
see  if invariance of the 1-body operators will persist.

In addition, we have made predictions~\cite{largen3} using the 1-body
operators plus the $c_3$ operator and can say which of the {\bf 70}-plet
decays are among the most sensitive to including the previous best-fit
value of the $c_3$ term. Here are some examples, good and bad [decay
amplitudes are all in (GeV)$^{-1/2}$]:
\begin{eqnarray}
A_{1/2}[\Delta^+(1620) \rightarrow \Delta\gamma] &=&
    \left\{ 
\begin{array}{ll}
   0.073 \pm 0.006 & \quad \mbox{{\rm 1-body},}  \\
   0.074 \pm 0.006 & \quad \mbox{{\rm 1-body} +} \ c_3,
\end{array}
    \right.  
                           \nonumber \\
A_{1/2}[p \, (1675) \rightarrow \Delta\gamma] &=&
    \left\{ 
\begin{array}{ll}
   -0.019 \pm 0.009 & \quad \mbox{{\rm 1-body},}  \\
   -0.060 \pm 0.013 & \quad \mbox{{\rm 1-body} +} \ c_3,
\end{array}
    \right.  
                           \nonumber \\
A_{5/2}[\Delta^+(1675) \rightarrow \Delta\gamma] &=&
    \left\{ 
\begin{array}{ll}
   -0.258 \pm 0.012 & \quad \mbox{{\rm 1-body},}  \\
   -0.337 \pm 0.020 & \quad \mbox{{\rm 1-body} +} \ c_3 .
\end{array}
    \right.
\end{eqnarray}

\vskip 3ex


\noindent{\bf 3. Mass Formula}


\vskip 2ex

\noindent  We have explained in the Introduction why we are going to discuss
a mass formula in the midst of this baryon decay talk: the derivation is
clear and illustrative of large $N_c$ methods, the original derivation is
dated, and we need the result.

We looking at a {\bf 56}-plet of baryons where the
spatial state, and so also the spin-flavor state, is totally symmetric. 
There are 56 totally symmetric 3-quark states that one can make from
$u_\uparrow$,
$u_\downarrow$,
$d_\uparrow$,
$d_\downarrow$, $s_\uparrow$, and $s_\downarrow$, where the arrows indicate the
spin projection.  The ground states form the {\bf 56}, and the
radially-excited  states form the {\bf 56$^\prime$}.  The states are the
$N$, $\Lambda$, $\Sigma$, $\Xi$, $\Delta$,
$\Sigma^*$, $\Xi^*$, and $\Omega$.

The mass operators for these states are built from 
the spin $S^i = \sum_\alpha \sigma^i_\alpha / 2$ (the sum is over the
quarks $\alpha$), the flavor operators $T^a = \sum_\alpha \tau^a_\alpha / 2$
(where the $\tau^a$ are a set of 3 $\times$ 3 matrices), and the SU(6)
operators
\begin{equation}
G^{ia} = \sum_\alpha {1\over 2} \sigma^i_\alpha \cdot {1\over 2} \tau^a_\alpha 
\ .
\end{equation}

Terms in mass operators must be rotationally symmetric, and
flavor symmetric to leading order.  Not all terms should be included. For
example, in symmetric states matrix elements of $T^2$ and $G^2$ are linearly
dependent on those of
$S^2$ and the unit operator~\cite{manrev}.

Flavor symmetry is not exact.  The mass of the strange quark allows non-flavor
symmetric terms in the effective mass operator, visible  as unsummed flavor
indices $a=8$ below.  The effective mass operator is 
\begin{eqnarray}
H_{\rm eff} &=& a_1 1 + {a_2\over N_c} S^2 + \epsilon a_3 T^8 + 
          {\epsilon\over N_c} a_4 S^i G^{i8}  \nonumber \\
         &+& {\epsilon \over N_c^2} a_5 S^2 T^8
          + {\epsilon^2 \over N_c} a_6 T^8 T^8 \nonumber \\
         &+& {\epsilon^2 \over N_c^2} a_7 T^8 S^i G^{i8}
          + {\epsilon^3 \over N_c^2} T^8 T^8 T^8 \ .
\end{eqnarray}

There is an $\epsilon$ for each violation of flavor symmetry, where $\epsilon
\approx 1/3$.  Also, a term that is a product of two or three operators comes
from an interaction that has at least one or two gluon exchanges, and the
strong coupling falls with number of colors as $g^2 \sim 1/N_c$ (A crucial
theorem is that no perturbation theory diagrams fall slower in $1/N_c$ than the
lowest order ones~\cite{manrev}).

Keeping the first four terms, taking the matrix elements, and
reorganizing leads to
\begin{eqnarray}
M &=& A + B N_s + C \left[I(I+1) - {1\over 4} N_s^2 \right] \nonumber \\
 &+& D S(S+1) ,
\end{eqnarray}
where $N_s$ is the number of strange  quarks.  This is the
G\"ursey-Radicati~\cite{GR} mass formula.  We use it to predict masses of 4
undiscovered members of the {\bf 56$'$}, given that 4 are known.

We can estimate the error of the formula.  The constants $a_i$ above are
nominally about 500 MeV.  The first term omitted is hence nominally of
order 500 MeV/$3^3$ or about 20 MeV, and this is the estimated error in
the G\"ursey-Radicati mass formula.  

\vskip 3ex


\noindent {\bf 4. The Decays {\bf 56$'$} $\rightarrow$ {\bf56} + Meson}

\vskip 2ex


\noindent  Four of the 8 states in the {\bf 56$'$} are undiscovered or
unconfirmed, and existing measurements have large uncertainty.
However, we anticipate new results soon from the CLAS detector at
CEBAF.  One member of the {\bf 56$'$} is the Roper or N(1440), whose
composition has been debated.  Might it be a $q^3g$
state~\cite{hybridroper}, a non-resonant cross-section
enhancement~\cite{dynamicalroper}, or just a 3-quark radial
excitation~\cite{isgurkarl1,glozmanriska}?  Recent lattice
calculations, which seem to find the negative-parity excited baryons
with about the right mass, still find the positive-parity excitation
at a fairly heavy mass~\cite{sasaki}.  Our predictions depend upon the
3-quark possibility.

We assume that only single-quark operators are needed.  Two-quark
operators were studied for decays of orbitally-excited
states~\cite{largen2}, and found unnecessary.  There is only one single
quark operator here, so
\begin{equation}
H_{\rm eff} \propto G^{ia} k^i \pi^a \ ,
\end{equation}
where $k^i$ is the meson 3-momentum and $\pi^a$ is a meson field operator.

One gets for the decay widths,
\begin{equation}
\Gamma = {M_f \over 6 \pi M_i} \, k^3 f(k)^2 \, 
     {\sum}  \  | \langle B_f | G_{ja} | B_i \rangle |^2 ,
\end{equation}
where $f(k)$  parameterizes the momentum dependence of the amplitude. For the 7
measured decays it is well fit by $f = (2.8 \pm 0.2)/k$.  With this in hand, we
can predict the widths for 22 decays. The detailed results are
in~\cite{largen3}.   The success of our predictions would bolster the view
of the Roper as a 3-quark state.

\vskip 3ex


\noindent{\bf 5. Conclusions}
\vskip 2ex


\noindent Here is a summary of our conclusions in list form:

\begin{enumerate}

\item For {\bf 70}-plet decays to $N\gamma$, single quark-operators are
not favored by large $N_c$, yet they suffice to explain the data.  Adding
extra operators does not materially improve the fit.  Why this should be so
remains a mystery.

\item  Decays of the {\bf 70}-plet to $\Delta\gamma$ add 24 new decay
channels to test large $N_c$ ideas or the approximation of using the
single-quark operators only.

\item  For {\bf 56$'$} decays (to $N\pi$), the mass spread
necessitated thinking about the momentum dependence of the matrix elements. We
got masses of the as yet unfound members of the
{\bf 56$'$} from the G\"ursey-Radicati mass formula---for which we have a
modern large
$N_c$ derivation.

\item Seven {\bf 56$'$} decays into baryon plus meson have been observed, the
remaining 22 being predicted.  We assumed the states, including the
Roper, were 3-quark states.  The success of the predictions could shed
light on whether this is true.

\end{enumerate}

\vskip 3ex


\noindent{\bf Acknowledgments}


\vskip 2ex

\noindent  I thank the conference organizers for their excellent work;
Chris Carone, Jos\'e Goity, and Rich Lebed for pleasant times
collaborating; and the  National Science Foundation for support under
Grant No.\ PHY-9900657.


\vskip 3ex

\noindent{\em Note added:} New data is coming in from JLab, and since the Workshop there has been an analysis~\cite{burkert02}, finding good results also for electroproduction (photon off-shell) with only single quark amplitudes.

\end{document}